\documentclass[twocolumn,aps,prl,superscriptaddress]{revtex4-2}

\usepackage{amsmath}
\usepackage{graphicx}
\usepackage{MnSymbol}
\usepackage{booktabs}
\usepackage{multirow}
\usepackage{rotating}

\begin{document}

\title{Sliding contact creates universal self-similar fractal surfaces}

\author{Ruibin~Xu}
\affiliation{State Key Laboratory of Solid Lubrication, Lanzhou Institute of Chemical Physics, Chinese Academy of Sciences, 730000 Lanzhou, China}
\affiliation{LTDS, CNRS UMR5513, Ecole Centrale de Lyon, 69134, Ecully, France}
\affiliation{Peter Gr\"unberg Institute (PGI-1), Forschungszentrum J\"ulich, 52425, J\"ulich, Germany}
\affiliation{MultiscaleConsulting, Wolfshovener str. 2, 52428 J\"ulich, Germany}

\author{Haohao~Ren}
\affiliation{State Key Laboratory of Solid Lubrication, Lanzhou Institute of Chemical Physics, Chinese Academy of Sciences, 730000 Lanzhou, China}

\author{Adriane~Clerc}
\affiliation{INSA Lyon, CNRS, LaMCoS, 69621, Villeurbanne, France}

\author{Guilhem~Mollon}
\affiliation{INSA Lyon, CNRS, LaMCoS, 69621, Villeurbanne, France}

\author{Wenbo~Sheng}
\email{shengwb@licp.cas.cn}
\affiliation{State Key Laboratory of Solid Lubrication, Lanzhou Institute of Chemical Physics, Chinese Academy of Sciences, 730000 Lanzhou, China}

\author{Feng~Zhou}
\email{zhouf@licp.cas.cn}
\affiliation{State Key Laboratory of Solid Lubrication, Lanzhou Institute of Chemical Physics, Chinese Academy of Sciences, 730000 Lanzhou, China}

\author{B.~N.~J.~Persson}
\email{b.persson@fz-juelich.de}
\affiliation{State Key Laboratory of Solid Lubrication, Lanzhou Institute of Chemical Physics, Chinese Academy of Sciences, 730000 Lanzhou, China}
\affiliation{Peter Gr\"unberg Institute (PGI-1), Forschungszentrum J\"ulich, 52425, J\"ulich, Germany}
\affiliation{MultiscaleConsulting, Wolfshovener str. 2, 52428 J\"ulich, Germany}

\begin{abstract}

Surface roughness evolves during sliding, a process known as run-in, yet the physical processes selecting the resulting topography remain unclear. Through experiments and analysis of measured surface topographies, we show that metals, rocks, and glasses develop universal self-similar roughness with a Hurst exponent $H\approx1$ at short wavelengths, while the long-wavelength roll-off remains material dependent. A minimal model of local material transfer associated with junction formation and rupture reproduces this short-wavelength scaling. We propose that competition between junction-mediated roughening and larger-scale plastic deformation and/or fracture sets the steady-state roll-off. These findings connect the diversity of run-in topographies to a common scaling structure and a physical picture of competing roughening and smoothing processes.
    
\end{abstract}

\maketitle

\vspace{0.3em}

\textbf{Introduction--}
All surfaces in natural and engineered systems are rough.
The real contact between asperities controls friction, wear, sealing, electric and thermal transport, and often governs performance and failure, from engineered interfaces to geological faults \cite{review1,APS0,APS1,APS2,APS4,Petrova2019,APS3,candela2012,n1,n2,n3,n4}.
Yet run-in topography is usually treated as an empirical outcome rather than a predictable design parameter, while the physical processes selecting it are frequently neglected.

Run-in is the transient period during which sliding surfaces reorganize before reaching a quasi-steady state of friction and wear \cite{Kragelsky1969,Blau2005,Blau2015}.
This steady state is not necessarily smooth: sliding may create new asperities as fast as it removes old ones.
Classic tribology introduced the notion of ``steady-state roughness'' \cite{Kragelsky1969,Blau2005,Blau2015}, and recent simulations have shown that 
self-affine roughness can emerge during both adhesive wear \cite{Milanese2019} and third-body
wear \cite{GarciaSuarez}.
Yet two basic questions remain open.
Do very different materials and loading histories evolve toward any common small-scale structure, or is the final roughness entirely system-specific?
And if common features exist, what microscopic processes select them?

Here we approach run-in as a problem of driven surface self-organization. We track the one-dimensional (1D) roughness power spectral density (PSD) $C_{\rm 1D}(q)$, which is proportional to the absolute square of the amplitude of the roughness component $h({\bf q})$ with wavevector ${\bf q}$ \cite{RP4}. We calculate $C_{\rm 1D}(q)$ from measured surface-height profiles to characterize topographical evolution and compare run-in surfaces across metals, rocks, glasses, and polymers. Our sliding experiments explore different initial roughnesses and nominal pressures under dry and boundary-lubricated conditions, with poly(methyl methacrylate) (PMMA) providing a polymer control. Despite large differences in material properties and starting state, the run-in surfaces of metals, rocks, and glasses exhibit a common short-wavelength regime where $C_{\rm 1D}(q)\sim q^{-\beta}$ with $\beta\approx3$, corresponding to a Hurst exponent $H=(\beta-1)/2\approx1$.
The approximate exponent is identified by comparing the short-wavelength power-law regions beyond the respective roll-offs with the reference scaling $C_{\rm 1D}(q)\propto q^{-3}$ on log-log plots.
This is the PSD of self-similar roughness, for which a magnified section of the roughness profile appears statistically similar to the original profile. Within the explored range of metal contacts, this short-wavelength scaling is insensitive to nominal pressure and to whether the surfaces are initially ``smooth'' (as-obtained) or ``rough'' (sandblasted). In contrast, the long-wavelength spectra remain strongly material dependent and exhibit distinct roll-offs. The common short-wavelength regime does not develop in boundary-lubricated steel or PMMA-metal contacts, consistent with a mechanism involving material transfer associated with the formation and rupture of strongly bonded microscopic contact regions, termed junctions.

High local stresses at asperity contacts can promote strong junction formation and material transfer \cite{Rabinowicz1951,Greenwood}. Related junction processes may also occur in brittle solids \cite{granite}, as discussed in the End Matter.

We interpret this combination of universality and specificity using a two-process picture.
At small scales, the repeated formation and rupture of microscopic junctions drive discrete material-transfer events that generate roughness.
As these local roughening events accumulate, their effect propagates toward larger length scales.
Because the same microscopic mechanism operates across scales, the resulting short-wavelength roughness naturally becomes self-similar, with $H = 1$.
At larger scales, plastic deformation and/or fracture act as smoothing processes that arrest this coarsening and set the roll-off wavenumber $q_{\rm r}$.
The competition between junction-driven roughening and large-scale smoothing therefore separates a universal small-scale regime from a material-dependent large-scale morphology.


\begin{figure*}[tbp]
\centering
\includegraphics[width=0.8\textwidth]{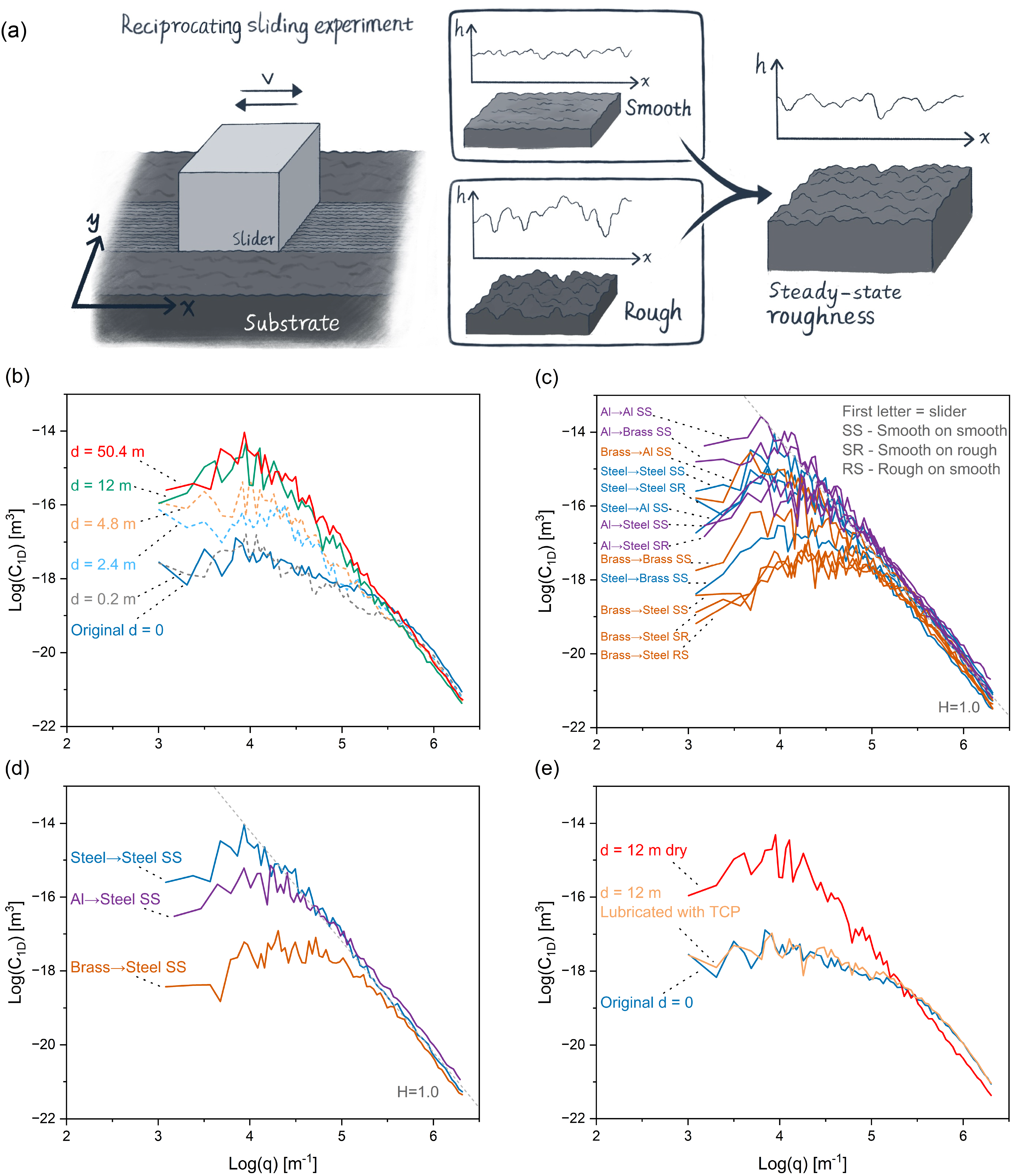}
\caption{Emergence of self-similar roughness during sliding. One-dimensional (1D) power spectral densities (PSDs) are measured perpendicular to sliding ($y$ direction). (a) Reciprocating sliding and topographical evolution. Here, $x$ is the sliding direction, $h$ the height, and $v$ the sliding speed. (b) PSD evolution for dry steel-steel contact with initially smooth surfaces. The spectra at sliding distances $d=12.0$ and $50.4$ m nearly coincide, indicating completion of run-in. The dashed line shows $C_{\rm 1D}(q)\propto q^{-3}$, corresponding to $H=1$. (c) Steady-state PSDs for dry metal contacts at $2.8$ MPa. SS, SR, and RS denote smooth-on-smooth, smooth-on-rough, and rough-on-smooth configurations, respectively, with the slider listed first. Al denotes aluminum. The spectra share approximately self-similar scaling beyond their material-dependent roll-offs. (d) Steel, brass, and aluminum sliders on smooth steel (SS), showing common short-wavelength scaling but distinct long-wavelength spectra. (e) Initial smooth steel (blue), after $12$ m with tricresyl phosphate (TCP) boundary lubrication (orange), and after $12$ m of dry sliding (red). TCP suppresses roughening and the development of the dry-contact $H=1$ regime.}

\label{combinedFig1}
\end{figure*}

\textbf{Surface roughness power spectra during run-in and at steady state--}
We performed linear reciprocating sliding tests on specimen pairs including
AISI 304 stainless steel, H59 brass, AA 6061 aluminum alloy, PMMA, and
granite, with controlled initial surface roughness. The as-obtained surfaces
are referred to as ``smooth'', while selected specimens were subsequently
sandblasted and are referred to as ``rough''. These terms refer to the
initial surface preparation and roughness amplitude, rather than to the Hurst exponent $H$.
In addition, we analyzed published topography data for glass samples slid on sandpapers \cite{plasticity} and for marble and quartzite pins tested under near-seismic conditions \cite{clerc}, using the same PSD procedure.
Unless stated otherwise, the PSDs shown in the main text are computed from profiles measured transverse to the sliding direction.

Fig.~\ref{combinedFig1}(b) shows the evolution of the surface PSD during sliding.
For dry sliding between smooth steel surfaces, the spectra show negligible changes beyond $12 \ {\rm m}$, indicating a steady state. The as-obtained surface already exhibits an $H\approx1$
region at the largest wavenumbers. The possible origin of this initial scaling is discussed below. In the dry metal-metal contact conditions studied, the steady state is reached after a sliding distance of order $10 \ {\rm m}$, which is much smaller than the imposed total sliding distance. 
In this steady state, the high-wavenumber regime exhibits power-law scaling $C_{\rm 1D}(q)\sim q^{-3}$, corresponding to self-affine fractal roughness with a Hurst exponent $H=1$, also known as self-similar fractal roughness.

The robustness of this short-wavelength regime is confirmed for a broad set of metallic systems. As summarized in Fig.~\ref{combinedFig1}(c), the steady-state 1D PSDs exhibit common self-similar scaling beyond their respective roll-offs. The transition into this regime is gradual and occurs at different wavenumbers for different material pairs. Within the explored range, the short-wavelength scaling is insensitive to the initial roughness configuration and nominal contact pressure (see also Fig.~S6). Fig.~\ref{combinedFig1}(d) further shows that different metal sliders on smooth steel share this scaling while retaining distinct roll-offs and long-wavelength spectra.

Independent AFM measurements of the brass surface after run-in in brass-brass contact agree well with the stylus spectrum over their overlapping wavenumber range and extend the same scaling toward shorter wavelengths (see Fig.~S3 in the Supplementary Information).

We also analyzed glass samples slid on sandpapers \cite{plasticity} and marble and quartzite pins tested under near-seismic conditions \cite{clerc}.
Despite fundamental differences in bonding, metallic versus covalent or ionic, and in sliding mode, linear reciprocation versus pin-on-disc, all these systems converge to the same short-wavelength self-affine scaling with $H = 1$ (see Fig.~S5(b)).
Thus, the universality of the high-$q$ regime is not limited to dry metal-metal contacts.

Boundary lubrication alters this behavior.
As shown in Fig.~\ref{combinedFig1}(e), in dry steel-steel contact the surface is already run-in after $12 \ {\rm m}$, whereas sliding in the presence of tricresyl phosphate (TCP) produces only negligible changes in the steel PSD, and the $H = 1$ regime does not develop. Varying the sliding speed and load does not change this outcome over a total sliding distance of $60 \ {\rm m}$ (Fig.~S7).
The same result is observed for PMMA-metal contacts, where the $H = 1$ regime is absent even after $64.8 \ {\rm m}$ (Fig.~S5(c)).

For systems that show an $H=1$ regime in the transverse direction, the same short-wavelength exponent is also observed along the sliding direction.
However, metals exhibit anisotropy between the two directions, with different PSD amplitudes but the same short-wavelength exponent, whereas granite and quartzite remain nearly isotropic.
The derivation and the material-specific results are given in the Supplementary Information.
We now turn to the origin of the universal $H = 1$ regime and then to the mechanism that sets the material-dependent roll-off.

\begin{figure*}[tbp]
\includegraphics[width=0.9\textwidth,angle=0]{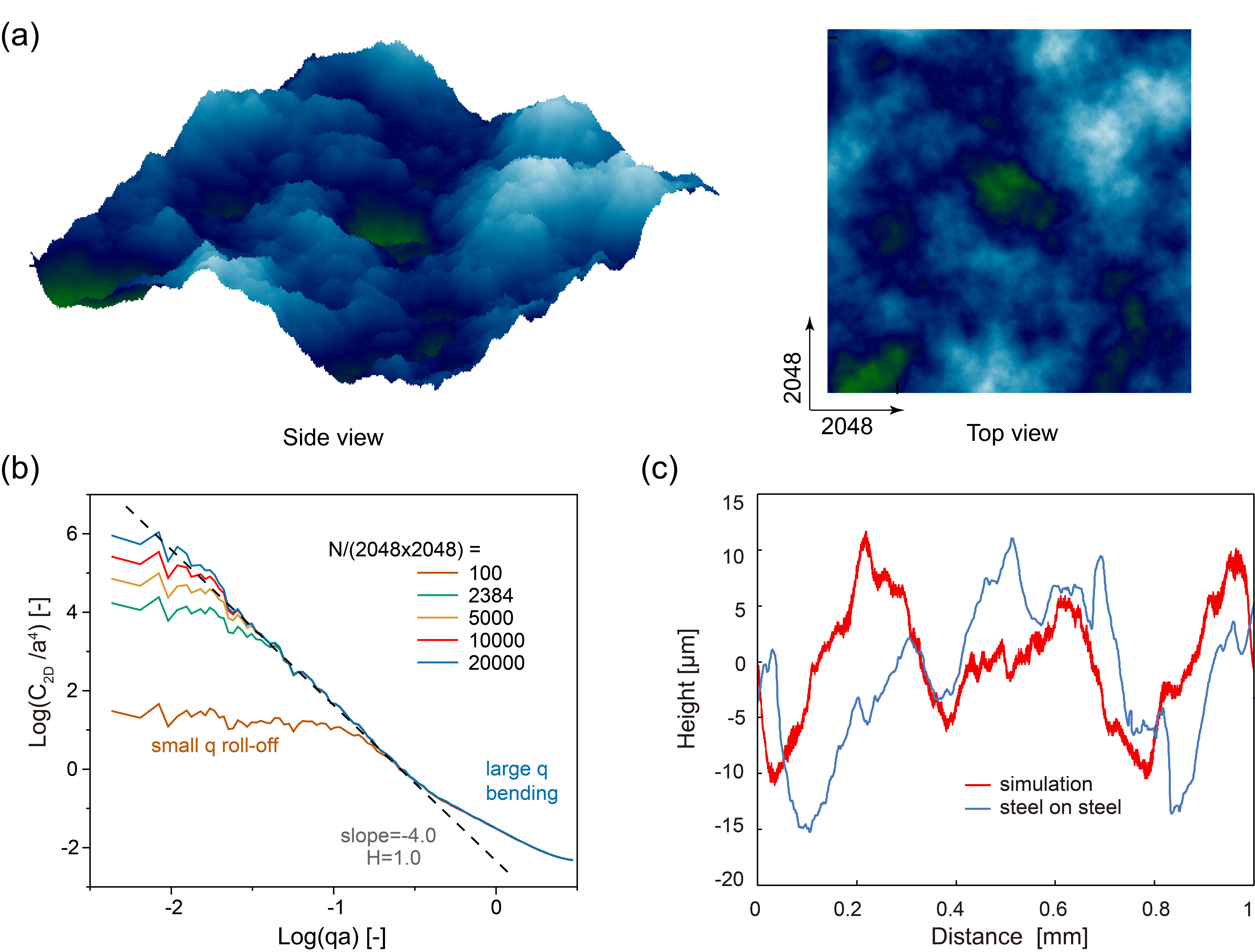}
\caption{Stochastic block-transfer model of self-similar roughness. (a) A $2048\times2048$ surface after $N\approx10^{11}$ block addition and removal steps (left, side view; right, top view). (b) Two-dimensional PSDs versus $qa$, where $q$ is the wavenumber and $a$ is the block size. The roll-off shifts toward smaller $q$ as $N$ increases. (c) Measured run-in steel profile (blue) and a line profile from the simulated surface in (a) (red). The simulated surface is statistically isotropic, so profiles along different directions have equivalent statistical properties.}
\label{combined}
\end{figure*}

\textbf{Origin of the $H = 1$ Hurst exponent--}
The steady-state PSDs exhibit self-similar scaling, meaning that the surface appears statistically unchanged upon magnification. One may ask how the roughness at one scale can ``know'' about that at another scale. One possible explanation is that roughness is generated through elementary steps at short length scales. As the number $N$ of elementary steps increases, roughness accumulates and propagates toward longer length scales. If a single mechanism generates roughness at all these scales, the roughness across scales will be correlated. This picture forms the basis of many models of surface growth and erosion \cite{Wolf,Binder} and leads to scaling laws that have been confirmed experimentally \cite{Stan}.


This picture may also explain why the as-obtained steel surface in Fig.~\ref{combinedFig1}(b) already exhibits an $H\approx1$ region at the largest wavenumbers before sliding. Plastic deformation, interfacial shear, and local slip during the production (sheet rolling) may have generated this scaling over a limited short-wavelength range. During subsequent sliding, the $H\approx1$ region progressively extends toward lower wavenumbers, corresponding to larger length scales. This evolution is consistent with the picture described above.

Here we argue that, at least for metal-metal contacts, the surface roughness results from the formation of cold-welded junctions and metal transfer \cite{JCPwear}. We assume that these junctions occur at short length scales and involve the transfer of characteristic volumes of metal between the two surfaces, termed ``blocks''.

Spatially uncorrelated block addition and removal produce a flat spectrum, rather than the observed $H=1$ scaling. We therefore use a simple model that mimics junction-mediated transfer through spatially correlated block addition and removal events generated by local asperity contacts. The full local rules and the uncorrelated $H=-1$ reference case are described in the End Matter. For a similar model, it was shown in Ref.~\cite{Persson} that starting from a flat surface and repeating the elementary steps produces self-similar roughness with $H=1$.

We study how the two-dimensional (2D) PSD $C_{\rm 2D}(q)$ depends on wavenumber after different sliding distances, represented here by different numbers of block addition and removal events. For a self-affine fractal surface, the 2D PSD scales as $C_{\rm 2D}(q)\sim q^{-2(1+H)}$, whereas the 1D PSD scales as $C_{\rm 1D}(q)\sim q^{-(1+2H)}$.

Fig.~\ref{combined}(a) shows a $2048 \times 2048$ surface generated by this process, and Fig.~\ref{combined}(b) presents the corresponding two-dimensional power spectra.
The roll-off region shifts toward smaller $q$ as $N$ increases, while the high-wavenumber regime exhibits the characteristic $H = 1$ scaling.
The simulated topography is also very similar to that observed for metals sliding on metals, as shown by the comparison with a measured steel line profile in Fig.~\ref{combined}(c).

This picture also explains the absence of the $H=1$ regime under boundary lubrication, which suppresses junction formation and material transfer. For TCP-lubricated steel-steel contacts, the PSD remains essentially unchanged after sliding. Similar suppression occurs in PMMA-metal contacts, where strong cold-weld-like junction formation is unlikely under the present conditions.

The model described above explains the emergence of the universal short-wavelength regime with $H = 1$.
However, the experimentally observed finite steady-state roll-off requires an additional mechanism, which we discuss next.

\begin{figure*}[tbp]
\centering
\includegraphics[width=\textwidth]{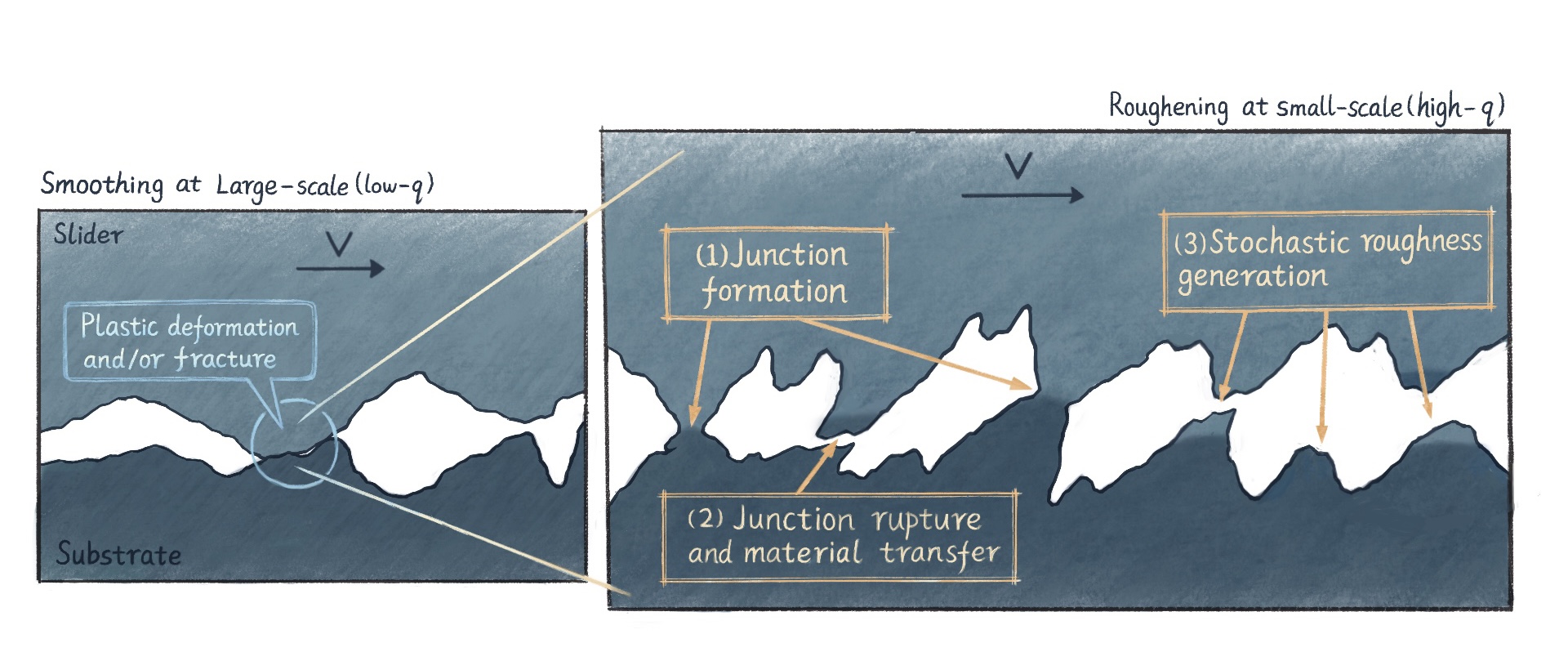}
\caption{Two-process picture of run-in. Larger-scale plastic deformation and/or fracture smooth asperities (left), while local junction formation and rupture generate roughness through material transfer (right). Their competition sets the roll-off wavenumber $q_{\rm r}$, separating short-wavelength self-similar scaling with $H=1$ from material-dependent long-wavelength morphology.}
\label{mechanism}
\end{figure*}

\textbf{Role of plastic deformations in setting the steady-state roll-off--}
The block-transfer model described above generates roughness whose roll-off shifts to smaller wavenumbers as the number of steps increases.
For an infinite system, and in the absence of smoothing, junction-mediated transfer would drive the long-wavelength roughness to increase without bound.
This is not what is observed experimentally.
Instead, the surfaces reach a finite steady-state roll-off at wavenumbers much larger than those set by the system size, showing that a competing smoothing mechanism must arrest the roughening.

We attribute this smoothing to plastic deformation and/or fracture of asperities.
If two solids made of the same material interacted only through short-ranged repulsion, that is, without any roughening process, then under repeated sliding the surfaces would become progressively smoother due to plastic deformation, until only elastic deformation occurred.
This smoothing process, often referred to as shakedown \cite{shake}, is time-dependent and requires a sliding distance that depends on the nominal contact pressure, the initial roughness, and the elastoplastic properties of the solids.

The roughening associated with the formation of cold-welded junctions is also time-dependent.
The roughening rate, i.e., the rate at which roughness propagates to longer length scales, depends on the probability of junction formation, which is influenced by the nature of any protective oxide or contamination film.
For a given smoothing rate, a higher roughening rate leads to a rougher surface, with longer-wavelength roughness and a larger root-mean-square (rms) amplitude. In this case, the roll-off will occur at larger length scales (smaller wavenumbers).
Thus, the roll-off wavenumber reflects a balance between junction-mediated roughening and plastic smoothing.

In summary, sliding produces common short-wavelength self-similar scaling with $H\approx1$ across metals, rocks, and glasses, while retaining material-dependent long-wavelength morphology. A minimal junction-transfer model reproduces this scaling, while competition with plastic deformation and/or fracture provides a physical interpretation of the finite steady-state roll-off (Fig.~\ref{mechanism}).

\textbf{Acknowledgements}
This work was supported by the Strategic Priority Research Program of the Chinese Academy of Sciences (Grant No.~XDB0470200).
We thank C.~Ma (LICP, China) for performing the AFM measurements and N.~S.~Kiselev (Forschungszentrum J\"ulich, Germany) for useful advice.

\textbf{Data and code availability}
The data supporting the findings of this study and the simulation code used for generating the model surfaces are available from the corresponding author upon reasonable request.

\clearpage
\textbf{End Matter}

\textbf{Minimal block-transfer model--}
Consider the sliding contact between two nominally flat solids with very small initial roughness. The substrate and slider are represented as cubic lattices of blocks of width $a$, and contact between an asperity of the slider and the substrate results in a block being either removed or added. We interpret these elementary steps as microscopic junction events, and the removal or addition of a block as the transfer of a small amount of material between the two surfaces.

High local stresses at asperity contacts during sliding can break contamination films and, for metals, oxide films, resulting in direct material contact, strong bond formation, and material transfer. Metallic transfer was observed by Rabinowicz and Tabor \cite{Rabinowicz1951,Greenwood} using autoradiographic measurements. Related cold-weld-like junctions can form even in brittle solids such as quartz \cite{granite} and may contribute to the high friction coefficient (of order 1) observed for granite sliding on granite.

The model includes only material transfer between the two surfaces, rather than wear processes. However, it could be extended to include the formation of wear particles by assigning a non-zero probability to the transfer of blocks out of the contact. As shown in Ref.~\cite{Persson}, this would also result in self-similar roughness. We also note that metallic wear begins only after a minimum sliding distance, during which sufficient material transfer has occurred, as discussed in Ref.~\cite{JCPwear}.

For identical materials, there is on average no net transfer of blocks between the slider and the substrate. For dissimilar metals, the initial transfer may be biased in one direction, for example from the substrate to the slider, but after some run-in the slider becomes covered by a thin transfer film of substrate material, and the transfer then becomes approximately symmetric \cite{JCPwear}.

We focus on the roughness evolution of one surface, while the opposite surface is not evolved explicitly.

Spatial correlations between transfer events are essential for the observed scaling. For comparison, spatially uncorrelated block addition and removal with a laterally uniform probability produce a wavevector-independent spectrum, $C_{\rm 2D}(q)=C_0$. Using the relation $C_{\rm 2D}(q)\sim q^{-2(1+H)}$ formally gives $H=-1$ for this reference process. The observed $H=1$ scaling therefore motivates the local selection rules described below.

First, consider block removal. If an asperity impacts at site $(i,j)$, chosen uniformly at random from all sites and located at $(x,y)=(ia,ja)$, then one of the blocks at $(i,j)$, $(i+1,j)$, $(i-1,j)$, $(i,j+1)$, or $(i,j-1)$ is removed. With probability $P$, the block with the smallest number of nearest neighbors is removed, since it is the most weakly bound. With probability $1-P$, the highest block is removed. If several candidates satisfy the same criterion, one is chosen randomly, except that block $(i,j)$ is selected whenever it belongs to the candidate set.

A schematic cross-sectional illustration of this local block-removal process is shown in Fig.~S8 of the Supplementary Information.

Next consider block addition. If an asperity impacts at site $(i,j)$,
with position $(x,y)=(ia,ja)$, one block is added to one of the sites
$(i,j)$, $(i+1,j)$, $(i-1,j)$, $(i,j+1)$, or $(i,j-1)$.
With probability $P$, the site with the largest number of nearest
neighbors is selected, since the added block is then most strongly bound.
With probability $1-P$, the lowest site is selected. If several candidates
satisfy the same criterion, one is chosen randomly, except that site
$(i,j)$ is selected whenever it belongs to the candidate set.

In Fig.~\ref{combined}, we used $P=0.5$. In the Supplementary Information, we present results for $P=0.3$, $0.5$, and $0.7$, all of which yield approximately the same exponent, $\beta\approx4$, in the power-law region of the 2D PSD, where $C_{\rm 2D}(q)\sim q^{-\beta}$. For a 2D PSD, this corresponds to a Hurst exponent $H=1$. Hence, the Hurst exponent $H$ does not depend on $P$ over the range studied. This is consistent with the theory of critical phenomena, in which critical exponents are generally insensitive to microscopic details (see Ref.~\cite{Stan} and the Supplementary Information).

Fig.~\ref{combined}(a) shows a $2048\times2048$ surface generated after $N\approx10^{11}$ steps of block addition and removal, and Fig.~\ref{combined}(b) presents the corresponding two-dimensional power spectra as a function of $qa$, where $q$ is the wavenumber and $a$ is the block size. The roll-off region in the log-log plot of $C_{\rm 2D}(q)$ shifts toward smaller $q$ as $N$ increases, while at high wavenumbers all spectra display a characteristic ``bending region'' where $C_{\rm 2D}(q)$ exceeds the extrapolated $H=1$ fit line. This bending region results from the finite size of the blocks added and removed and is likely to occur also in the systems studied here, where it would be related to the size of the cold-welded junctions, but at wavenumbers beyond the highest values probed experimentally.

As an example, Fig.~\ref{combined}(c) compares the measured line profile for the steel surface after run-in with a line scan extracted from the simulated $2048 \times 2048$ surface with $a = 0.5 \ {\rm \mu m}$ after $N \approx 10^{11}$ steps.
The theoretical curve appears thicker than the measured one because of the enhanced short-wavelength roughness arising from the high-wavenumber bending region in the power spectrum. Such bending may also occur for the steel surface at still larger wavenumbers, beyond the experimentally accessible range of the present stylus measurements.

The model also provides a way to estimate the characteristic size of the elementary transfer events. The stylus spectra exhibit self-similar scaling down to wavelengths of order $1~\mu{\rm m}$, suggesting that the elementary transfer size lies below this scale. Additional AFM measurements on the brass surface extend the same scaling to shorter wavelengths without revealing a clear crossover associated with the elementary transfer size (Fig.~S3).
This suggests that the cold-welded junctions responsible for the elementary transfer events have lateral dimensions below about $0.1~\mu{\rm m}$ for the brass-brass contact studied here.
This scale is much smaller than the typical wear particles observed experimentally, which suggests that the wear particles, often flake-like, consist mainly of lateral aggregates of transferred material rather than single elementary transfer units \cite{Rabinowicz1951,Greenwood,JCPwear}.

\textbf{Finite roll-off and smoothing--}
In the model presented above, there is no plastic smoothing, only the roughening process.
For an infinitely large system, the block-transfer model generates roughness with an rms amplitude that increases without bound as the number of steps $N \to \infty$.
More precisely, the model predicts a roll-off region that continuously shifts to smaller wavenumbers, corresponding to longer length scales, as $N$ increases, with $q_{\rm r} \sim N^{-1/2}$.
For a finite system of linear size $L$, this process would terminate only when the roll-off wavenumber reaches $q_{\rm r} \sim \pi / L$.

However, this is not what is observed experimentally.
For the present system, $L \approx 1 \ {\rm cm}$ gives $\pi / L \approx 300 \ {\rm m}^{-1}$, whereas the measured steady-state roll-off occurs at much larger wavenumbers, of order $10^{4} \ {\rm m}^{-1}$.
Thus, the experimentally observed finite steady-state roll-off cannot be explained by junction-mediated roughening alone.
A competing smoothing mechanism must arrest the roughening before the roll-off reaches the system size.
The smoothing occurs by plastic flow and does not involve material transfer between the two surfaces. Thus, when asperities at the largest length scale make contact, we envisage them deforming (flattening) plastically.

The steady-state roll-off therefore reflects a balance between junction-mediated roughening and plastic smoothing.
This picture might also explain why brass-brass exhibits a smoother long-wavelength spectrum than steel-steel or aluminum-aluminum.
If the probability of block transfer is smaller for brass-brass, then the roughening rate will be lower.
If the smoothing rates are similar, the final topography for brass-brass will therefore be smoother, consistent with the observed results.
The same trend would arise if the characteristic block size $a$ were smaller for brass than for steel.
Thus, the physical origin of why some material combinations result in smoother surfaces could be either a lower probability of junction formation and material transfer, or a smaller effective transfer-unit size, both of which reduce the roughening rate at a given smoothing rate. Alternatively, a higher smoothing rate could also produce a smoother long-wavelength spectrum at a given roughening rate. The present experiments do not separately determine these rates.

The present block-transfer model results in roughness with isotropic statistical properties,
and therefore does not explicitly describe the directional anisotropy observed for some of the
sliding surfaces. Such anisotropy could be incorporated by allowing the
elementary transfer events to become direction-dependent. Within our model,
this direction dependence may originate from the stretching that microscopic
junctions sustain along the sliding direction before rupture. Greater
junction elongation would then produce stronger anisotropy. This is
consistent with our observations if such stretching is greatest for brass,
smaller for aluminum and steel, and negligible for granite and quartzite,
which remain nearly isotropic. Since all the studied surfaces exhibit the
same Hurst exponent $H=1$, our results suggest that the anisotropy resulting
from such stretching does not change the universality class of the roughening
process (see Supplementary Information). 
However, how the anisotropy depends on the plastic properties of the materials (ductility) remains an extremely interesting question for future studies.

Both the smoothing and the roughness anisotropy in the self-similar regime depend on the plastic flow properties of the solids, such as ductility. Addressing these questions quantitatively represents an important direction for future research.


\end{document}